\documentclass[prl,onecolumn,showpacs,floatfix,showkeys]{revtex4} 

\usepackage{latexsym,amssymb,amsmath,amsfonts}
\usepackage{graphicx}
\usepackage{bm}
\newcommand{\beq}{\begin{equation}}
\newcommand{\eeq}{\end{equation}}
\newcommand{\bc}{\begin{center}}
\newcommand{\ec}{\end{center}}
\newcommand{\eeqa}{\end{eqnarray}}
\newcommand{\beqa}{\begin{eqnarray}}
\newcommand{\no}{\noindent}
\newcommand{\pa}{\partial}

\newcommand{\ra}{\rightarrow}

\newcommand{\al}{\alpha}
\newcommand{\be}{\beta}
\newcommand{\ga}{\gamma}

\newcommand{\De}{\Delta}

\newcommand{\la}{\lambda}

\newcommand{\si}{\sigma}

\newcommand{\ta}{\tau}

\newcommand{\om}{\omega}
\newcommand{\ed}{\end{document} }

\begin{document}

\title{Electrodynamics with radiation reaction}
\author{Richard T. Hammond}

\email{rhammond@email.unc.edu }
\affiliation{Department of Physics\\
University of North Carolina at Chapel Hill\\
Chapel Hill, North Carolina and\\
Army Research Office\\
Research Triangle Park, North Carolina}

\date{\today}


\begin{abstract}
The self force of electrodynamics is derived from a scalar field. The resulting equation of motion is free of all of the problems that plague the Lorentz Abraham Dirac equation. The age-old problem of a particle in a constant field is solved and the solution has intuitive appeal.
\end{abstract}

\maketitle

\section{Introduction}

It is well-known that a charged particle emits electromagnetic energy when it is accelerated. It is also well-known that this produces a self force that acts on the particle being accelerated. What is not well-known is the equation of motion of such a particle. Although there have been many other attempts, which are described in a recent review,\cite{hammondreview} no single equation as yet has gained a general consensus as to its authenticity and correctness.

Radiation reaction was discovered by Abraham,\cite{abraham} extended by Lorentz\cite{lorentz}
and rederived by Dirac,\cite{dirac} so that the LAD equation is often the starting point. The LAD equation is

\begin{equation}\label{lad}
\frac{dv^\mu}{d\tau}=\frac e{mc} F^{\mu\si}v_\si +\tau_0\left( \ddot v^\mu 
+\frac{v^\mu}{c^2} \dot v_\si \dot v^\si\right)
\end{equation}

\no where the overdot represents differentiation with respect to proper time, $\ta$.
This equation is generally assumed to be invalid since it leads to the infamous runaway solutions. For example, in the low velocity limit this gives $v=v_0e^{t/\ta_0}$ where $\tau_0=2e^2/3mc^3$ which, for an electron, is somewhat less than $10^{-23}$ s. This absurd solution has the electron reaching the speed of light in practically no time, even though no forces act on it. To counter this, one standard trick is to reduce the equation to a second order differential equation using an integrating factor, but the solution to the resulting equation violates causality.\cite{jackson}

Landau and Lifshitz sidestep the pitfalls by using an approximate form of the LAD equation.\cite{landau} It was shown that the LL equation could be derived by using an asymptotic expansion of the velocity in terms of $\tau_0$.\cite{hammondasym} The biggest problem with their result is that it is only valid for small radiation effects. For intensities of $10^{22}$ W cm$^{-2}$ currently achieved, and higher intensities sitting on the horizon, it has been shown that the self force is no longer small.\cite{hammondpra} Another equation of note is the Ford O'Connell equation.\cite{ford} 

All three of these equations suffer a peculiarity: They predict that a charged particle in a constant electric field is unaffected by the self force even though it radiates energy all the while it accelerates. This issue has caused a longstanding debate, with early (and incorrect) suggestions that the particle does not radiate at all, to the somewhat conventional, but grudging, acceptance of a mysterious induction field that accounts for the energy discrepancy.
Due to these difficulties, many authors have studied this problem resulting in a wide range of attempts that may be found in the literature.\cite{hammondreview} Many are based on a series, or improvable assumptions that the radiation reaction is a small effect, or have other questionable assumptions.

 None of these is accepted by the physics community as a whole to be the correct equation. The LAD equation leads to unphysical effects, and therefore all equations derived from it or equivalent to it lack credibility. To emphasize this point, a perpetual motion  is described.
 
\section{Perpetual Motion Machine}

The idea is based on the LAD prediction that a charged particle in a uniform, constant electric field, is not affected by the self force. ``Not affected'' means precisely this: Consider the solution to the equation of motion if the self force is ignored, and call this solution $v_i^\mu$. For example, if the particle starts off from rest and the electric field points in the $x$ direction, then the solution is $v_i^1=c\ \mbox{sinh}\Omega \ta$ (details are below). This solution conserves energy: The kinetic energy of the particle equals the work done by the external field. Now suppose we turn on the self force.  In this case the particle radiates away energy at the rate $P=m\ta_0c^2\Omega^2$, but
  $v_i^1$ {\em is still the solution} to the equations of motion. This being the case we still have, ``The kinetic energy of the particle equals the work done by the external field,'' and so we have  extra energy  that is radiated and may be used. For example if we can devise a cyclic process with a constant field, then the radiated energy can be used for other purposes, or sold to the power companies.

The perpetual motion machine does exactly this.
The machine is an annulus with a charge. The radius of the outer sphere is $a$ and the inner radius is $b$, and it is endowed with a charge density $\rho=3 Q/4\pi b^2 r$ where $Q=2b^2Q_T /3(a^2-b^2)$ where $Q_T$ is the total charge and is assumed to be positive. This gives $E=Q/b^2$, a constant, in the annulus and zero for $r<a$. Now we drill a straight thin tunnel completely through, passing through the center, and release a negatively charged particle from rest at the surface, at the tunnel.

The solution to (\ref{lad}) is, for the magnitude of the velocity $v$ (in the following the subscripts denote the region,

\beqa
v_1/c=\mbox{sinh}(\Omega \ta)\ \ \ \ \ a>r>b\\ \nonumber
v_2/c=\mbox{sinh}(\Omega \ta_1)\ \ \ \ |r|<a\\ \nonumber
v_3/c=-\left(\mbox{cosh}\Omega\ta_2\mbox{cosh}\Omega\ta_1+ \mbox{sinh}\Omega\ta_2\mbox{sinh}\Omega\ta_1
\right)\mbox{sinh}(\Omega \ta)\\ \nonumber
+\left(\mbox{cosh}\Omega\ta_2\mbox{sinh}\Omega\ta_1+ \mbox{sinh}\Omega\ta_2\mbox{cosh}\Omega\ta_1
\right)\mbox{cosh}(\Omega \ta)\\ \nonumber
a>r>b\\ \nonumber
\eeqa

\no where $\ta_1$ is the time when the particle first reaches $r=a$ and is given by

\beq
\ta_1=\frac{\mbox{arccosh}\left( \frac{(a-b)\Omega}{c} +1  \right)}{\Omega}
,\eeq

\no $v_2$ is the coasting velocity in the field free region and
$\ta_2=\ta_1+2a/v_2$ is the time when the particle first re-enters the annulus,
and $\Omega= qE/mc$.

At time $\ta_3$ the particle reaches the opposite side of the annulus, $v_3(\ta_3)=0$, and the motions continues {\it ad infinitum}. After one complete cycle the machine emits an energy $W$,

\beq 
W=4mc^2\ta_0(a-b)\mbox{sinh}\Omega\ta_1
,\eeq

\no the particle is precisely where it was at the beginning, and it cycles on forever, all the while giving up usable energy from the radiated power.

Of course, even if this theory (LAD equation) were correct, the perpetual motion machine is unrealistic in that it consists of perfectly uniform charge and step function boundaries (although by making the machine large enough these effects can be made small, by comparison). The point is to show the LAD equation, for the constant force problem, leads to an unphysical result.

This is not the only problem with the self force: The Schott energy comes from the term that is a third order differential. This is the term that leads to runaway solutions mentioned above, solutions that violate conservation of energy. A particle with some arbitrary initial velocity is accelerated to nearly the speed of light in a very short time. The final energy minus the original energy is large and positive, larger than the original energy, so clearly energy is violated. If the energy were to be stored in an induction field to balance this and conserve energy, we face the a problem: the stored energy would have to be negative. This is a problem because the energy of an electromagnetic field is positive definite.

In the following, a simple derivation is presented of the equation of motion, with the self force, that contains none of the difficulties mentioned above.

\section{The self force}

We seek to find the self force, $f^\si$, so that

\beq
m\dot v^\si=e F^{\si\mu}v_\mu +f^\si
.\eeq

\no We know the power radiated by an accelerated particle, $P=-m\ta_0\dot v_\si \dot v^\si$, is a scalar, and since proper time $d\ta$ is also a scalar, we know 
 $d\la\equiv Pd\ta$ is a scalar and moreover, since $d\la =\la,_\si dx^\si$ we know $\la,_\si$ is a vector. From its definition we see that $\la$ is associated with the energy of the radiation field.

In general, when a particle is subject to a force, that force is defined as the gradient of an energy (the potential energy), and taking this as a clue, we assume the self force may be constructed from terms linear in the derivatives of $\la$. This leaves

\beq\label{f1}
f^\si=\al \la^{,\si}+\be \dot \la v^\si
\eeq

\no where $\al$ and $\be$ are constants to be determined. A term like $\la\dot v^\si$ is excluded because we only consider forces that arise from derivatives of $\la$, as explained above. It can also be shown that such a term leads to unphysical solutions in the case of a particle in a constant magnetic field.
 Higher derivative terms in the velocity are ruled out because they give rise to unphysical processes like runaway solutions. 

Using the conditions $v_\si\dot v^\si=0$ we find $\be=-\al/c^2$, so that
the equation of motion is

\beq\label{al}
m\dot v^\si=e F^{\si\mu}v_\mu +\al( \la^{,\si}-\frac{1}{c^2} \dot \la v^\si)
\eeq

\no in terms of one unknown constant. It is Lorentz invariant from its construction and, naturally, valid in any reference frame.
One may note the self force may be written as
$f^\si=\al( \la^{,\si}-\la,_\mu v^\mu v^\si/c^2)$. Since $v^\mu/c \sim \ga$, for large $\ga$ we may neglect $ \la^{,\si}$ in the self force.

To find $\al$ we consider the lab frame, the frame in which $F^{\si\mu}$ and $v_\mu$ are given. We also consider the limit that $\ga>>1$. Integrating the zero component of (\ref{al}) with respect to the proper time gives

\beq\label{approx}
m(\ga-\ga_0)=W_F-\al\int P dt
\eeq

\no where $\ga_0$ is the initial value of $\ga$ and $W_F$ is the work done by the external field. The energy radiated--as measured in the lab frame-- is $\int P dt$ since the lab frame measures $t$. If we assume  the change in kinetic energy of the particle equals the work done by the external field minus the energy radiated in this limit, we have $\al=1$.

And so, we have the final Lorentz invariant equation of motion with the self force,

\beq\label{eqm}
m\dot v^\si=e F^{\si\mu}v_\mu + \la^{,\si}-\frac{1}{c^2} \dot \la v^\si
\eeq

\no and, from before,

\beq
\dot\la=-m\ta_0\dot v_\si v^\si
.\eeq

However, from virtually every approach to this problem we find that (\ref{approx}) is not valid, and here it is only an approximation. To see the exact relations let us consider the general case,

\beq\label{eqm0}
m\dot v^0=e F^{0\mu}v_\mu + \la^{,0}-\frac{1}{c^2} \dot \la v^0
,\eeq

\no and once again integrate each term with respect to proper time $d\ta$ in the lab frame, 

\beq\label{energy}
m(\ga-\ga_o)=\int {\bm F} \cdot d{\bm x} - \int P dt
+\int \pa_t \la d\ta
.\eeq

This equation reads, the change in energy of the particle equals the work done by the external force minus the energy radiated away plus the last term.
In the LAD type equations this corresponds to the infamous Schott term, the term responsible for runaway solutions. In the present case this term finds a more benign interpretation and may be viewed as the work done by the radiation field on the particle.
 The radiation field produces a force, and as the particle moves there is the resulting term: force times distance, and so, it is natural to assume this is the work done by the radiation force. 
 In words we may say,

\begin{align} 
  \mbox{(change in kinetic energy)} =\\ \nonumber \mbox{(work done by total field)} - \mbox{(energy radiated)}\nonumber 
 \end{align}
 
\no or, equivalently 
 \begin{align} \label{energy2}
  \mbox{(change in kinetic energy)} =\\ \nonumber \mbox{(work done by external field)}\\ \nonumber
  +\mbox{(work done by radiation field)}
     - \mbox{(energy radiated)}\nonumber 
 \end{align}
 
\no where total field means external plus radiation field.

This is a very pleasing result. It has long been known that ``some of the energy goes missing.'' In other words, we have a physical interpretation of each term except the Schott term, in which case it is usually argued the energy is stored in a mysterious induction field. In the present case we see it corresponds to the work done by the self force, a term that should appear in an energy balance equation in any event.

  In the low velocity limit (\ref{eqm}) becomes
 
\beq
m\dot{\bm v}= q{\bm E}-{\bm\nabla \la}
.\eeq

\no In this limit we see that our original ideas are on track, and that $\la$ is a like a potential energy of the radiation field.

\section{constant acceleration: The Ephemeral perpetual motion machine}

As an example, we may solve another infamous problem, the problem of a charged particle in a constant, uniform field. This problem has been a raging storm. Perhaps the first clouds were formed by Pauli, who incorrectly claimed that in such fields particles do not radiate.
Later, the principle of equivalence was incorrectly used and, although it is now recognized that the particle does radiate, the correct equations of motion have been as elusive as ever.

 The essence of the difficulty, using the LAD equation, is that the solution to the equation of motion with the self force is the same as that without the self force, even though the particle is radiating all the time. This absurd result gives rise to the perpetual motion machine described above.

We now show that we can provide sensible solutions to (\ref{eqm}) for a constant electric field,

\beq\label{const0}
\dot v^0=\Omega v^1-\frac{\dot \la}{mc^2}v^0+\la,_0/m
\eeq

\beq\label{const1}
\dot v^1=\Omega v^0-\frac{\dot \la}{mc^2}v^1-\la,_1/m
\eeq

\no where $\Omega=eE/mc$. To solve these we use an asymptotic series approach in terms of $\ta_0$ where the zero order solutions,
$v^0=c\ \mbox{cosh}(\Omega\ta)$ and $v^1=c\ \mbox{sinh}(\Omega\ta)$, are used to compute the self force, and then these are used to solve (\ref{const0}) and (\ref{const1}). In this case the power is $\dot\la=P=m\ta_0c^2\Omega^2$, $\la,_0=P/\ga c$ and $\la,_n=0$. To simplify things, we will adopt dimensionless units, so the solution to (\ref{const0}) and (\ref{const1}) becomes, using $b=\Omega\ta_0$

\beq
v^1=\mbox{sinh}\ta -b \mbox{cosh}\ta\left(
\ta +\ln(e^{-\ta}\mbox{cosh}\ta)
\right)
,\eeq

\no which is valid for $b<<1$. It may be noted that, with this, the $v1 < \mbox{sinh}\ta$, which tells us the velocity that includes the self force is less than the velocity that is found excluding the self force ($\mbox{sinh}\ta$).

With this solution we have a sensible energy relation.
Let us consider the kinetic energy, $K=v^0-1$. To keep things simple we continue using the non-dimensional solution and work on a per unit mass basis.  Now let us write out the zero component of the equation of motion,

\beq
\dot v^0= v^1+ P/v^0 -Pv^0
\eeq

\no and integrate each term with respect to the proper time.  We define these integrals as

\beq 
W_F=\int v^1d\ta
\eeq

\no which is the work done by the external field,

\beq
W_S=\int \frac{P}{v^0}d\ta
,\eeq

\no which we interpret as the work done by the self force,

\no and
\beq
W_R=\int Pv^0d\ta
,\eeq

\no the energy radiated away.
In terms of these quantities (\ref{energy2}) becomes

\beq\label{energy3}
K=W_F+W_S-W_R
.\eeq

The  integrals give

\beq
K=\mbox{cosh}\ta-1\\ 
-b \ \mbox{sinh}\ta\left(
\ta +\ln(e^{-\ta}\mbox{cosh}\ta)
\right)
\eeq

\beq
W_S=2b \arctan\left(\mbox{tanh}(\frac \ta 2)\right)
\eeq

\beq
W_R=b\ \mbox{sinh}\ta
\eeq

\beqa
W_F=\mbox{cosh}\ta-1
-2b\arctan\left(\mbox{tanh}\frac \ta 2\right)
\\ \nonumber
-b\ \mbox{sinh}\ta\left(-1
+\ta+\ln\left(e^{-\ta}\mbox{cosh}\ta\right)\right)
.\eeqa

\no We see that (\ref{energy2}) holds explicitly, and moreover that $|W_R|>|W_S|$. This tells us that the kinetic energy is less than the work done by the external field, or, the velocity in this case is less then the LAD result. Thus the particle's amplitude of oscillation dwindles, and we see what we knew along, there is no perpetual motion machine.

\section{High energy}
The self force is a small effect in the NR limit, but becomes important as the energy becomes large. For example, as noted above, at currently producible laser intensities of $10^{22}$ W cm$^{-2}$ the self force is important, and will become more so, if not dominate, as higher intensities are reached.
The high energy limit is characterized by having the relativistic factor $\ga>>1$.
In this limiting case we find the self force becomes
 $f^\mu \ra -v^\mu P/c^2$.

In Ref. I numerical solutions were given for the high energy case. It was shown how energy was conserved and possible ways to experimentally verify the self force were discussed. Now we shall consider an analytical solution, a solution that we might have said ``must exist,'' but has never in fact shown to exist.

The problem is simply this: We consider a charged particle in a plane electromagnetic wave.
It is generally agreed that acceleration by a plane wave is not possible due to the Lawson Woodward theorem, even though many spirited debates have arisen over this.\cite{lawson} Here we show that acceleration by a plane wave is not only possible, but agrees with the well-known Compton scattering result.

In its simplest form the LW theorem states that a plane electromagnetic wave, pulsed or infinite, cannot impart a net energy to a charged particle. This theorem also assumes there are no radiation reaction effects. In other words, if a charged particle is irradiated by a plane wave, the change in the kinetic energy of the particle, measured from the time before the plane wave hits to after it leaves, is zero. At first glance this certainly seems wrong. For example consider a plane electromagnetic wave incident on an electron. The time average Poynting vector,
$S=cE^2/8\pi$, gives rise to the radiation pressure ${\cal P}=S/c$. If we multiply the pressure times the area we get a net force in the direction of the wave. Using the Thompson cross section for the area we find the force $ F$,

\begin{equation}\label{ff}
F=\si_T{\cal P}=\frac{a_c^2 E^2}{3}
,\end{equation}

\no which is also known as the Eddington force. The quantity $\sigma_T$ is $8\pi a_c^2/3$ where $a= e^2/mc^2$ and $m$ is the mass of the particle. Of course, there is no guarantee this mix of classical and quantum physics should give an exact answer, but it does indicate that the particle should be accelerated. Extending a single electron to a sheet of electrons, as in metallic film (or solar sail), we know there is indeed a force due to the radiation pressure. In fact, this question has been debated recently.\cite{rothman}

In contrast to these expectations, a classical calculation confirms the Lawson Woodward theorem. It is known the particle moves with the ``figure eight'' pattern and gains no net energy from the wave.\cite{hammondreview}
However, as noted, the Lawson Woodward theorem assumes radiation reaction is absent. It was noted a while ago that the existence of the self force will, in fact,  cause a particle to accelerate.\cite{fradkin}

As an example, consider the same plane wave as used above.
Although these equations are valid for arbitrarily high fields, let us consider an asymptotic expansion $v^\si= {_0v}^\si+ \ta_0 ( {_1}v^\si)$. The solution to ${\cal O} (\ta_0^0)$ can be found in the review \cite{hammondreview}, and these may be used to find find $v^\si$ to ${\cal O} (\ta_0)$. If we take the time average of this result for the $z$ component of the velocity we find

\begin{equation}\label{pt}
	\frac{<v^3>}{c} =\ta_0\om^2 a^4\ta^2/8
\end{equation}

\no  which shows the acceleration is in the direction of propagation of the beam in terms of the proper time.

The relation between lab time (the frame in which the electric field is given by that stated above) and proper time may be found from in \cite{hammondreview}, so (\ref{pt}) becomes, after averaging $<v^3/c>=\ta_0\Omega^2t/4$. Since the result is linear in time, suppose we assume there exists an equivalent force $F$ such that
$F=ma$. This may not be considered the true force acting on the particle because the problem has been solved in the relativistic regime,  but that is unimportant. The result is correct and now we are looking for a heuristic way to describe it. So, continuing, and using $a= dv/dt$ (average is now implied) we find $F=a_c^2E^2/6$, exactly half of that found by using the radiation pressure. This is in accord with the well-known result that a quantum mechanical cross section is usually twice the classical result. When using the Thompson cross section we are adopting the result of a quantum mechanical calculation, so the factor of 2 is not surprising. 

This result may also be written in terms of $\ga_F$, the value of $\ga$ after the particle has been in the field for a time $t$, and the intensity $I$,

\beq\label{ft}
\ga_F=\frac12\frac{\si_T}{mc^2}It
\eeq

\no a very simple result with intuitive appeal. It tells us that the energy gained by the particle per second is (one half) the intensity of the source times the area of the particle.
 The result agrees with the heuristic notion of radiation pressure pushing the particle. Although  Lawson Woodward  outlaws such motion, it is based upon the assumption that radiation reaction is not present, so the theorem is not violated--it does not apply. A harbinger of this result was found numerically already in \cite{hammondpra} where, for pulses of a few wavelengths, it was shown that with radiation reaction taken into account, the particle was accelerated in the forward direction.

Although this calculation was done for a monochromatic field, we see that the result (\ref{ft}) is independent of the frequency. Now, for an arbitrary spectrum we add the fields, but each frequency will produce the above result. Thus we may simply add the intensities and use the total intensity in (\ref{ft}).

We may now explore some consequences of this result. The only proviso is that the above result is valid $\ga>>1$, so we are dealing with relativistic particles. Let us consider the problem of UHECRs, ultra-high energy cosmic rays. Although there is no universally agreed upon mechanism that gives rise to the energies of UHECRs, it is agreed that it must come from a astronomical event of enormous energy (unless we consider top down events such as the decay of some exotic particle). So let us consider a gamma ray burster (GRB) of luminosity $10^{52}$ erg/sec\cite{ramirez} and assume it results from the collapse of a 35 solar mass star.\cite{vanmarle} We may obtain an estimate of the intensity by dividing the luminosity by the area of the event horizon.

Suppose we consider an electron that experiences the average radiation force derived above. The change in energy the electron experiences, from the surface of the star to infinity, is $\De U =\ta_0 e^2L/2mc^2r_h$. We may attribute this to a potential difference $V=\De U/e$. Now, as these electrons are blown away charge builds up on the star until it creates a critical field that stops further electrons from escaping. Equating this attractive force to the outward force we find, at the surface the field is $E=\ta_0 eL/2mc^2r^2_h$ which is about $5\times10^{15}$ V/m, about an order of magnitude less than the Schwinger value for pair creation.
Now, a proton will experience an acceleration due to the potential difference, which is $eV$. For a proton this corresponds to the relativistic factor
$\ga\approx 5\times10^{11}$, which corresponds to the most energetic UHECRs measured on earth.

In order to properly understand UHECRs one must be able to explain the spectrum, the number of events per energy range, and hopefully explain both the knee and the ankle in these curves. The point made here is that electromagnetic waves of sufficient intensity {\it may accelerate protons} to UHECR energies.

\section{Summary}

This article has shown three major developments  First, we give a heuristic derivation of the self force. The second accomplishment is the solution to the age-old problem of constant acceleration. The final accomplishment is the application to the high energy regime to derive an analytical result. This gives the vacuum acceleration of a charged particle in a strong electromagnetic field. This has never been done. Averaging out the fluctuations, it is shown that the result is in accord with the intuitive notion that the Poynting vector should impart a momentum to a charged particle. A possible application of this result is speculated upon.

\ed